# Suppressed charge recombination in hematite photoanode *via* protonation and annealing

Wenping Si,*,†,‡ Fatima Haydous,† Ugljesa Babic,§ Daniele Pergolesi,† and Thomas Lippert*,†,⊥

† Thin Films and Interfaces, Research with Neutrons and Muons Department, Paul Scherrer Institut, 5232 Villigen PSI, Switzerland

‡Key Laboratory of Advanced Ceramics and Machining Technology, Ministry of Education, School of Materials Science and, Engineering, Tianjin University, Tianjin 300072, P.R. China

§Electrochemistry Laboratory, Paul Scherrer Institut, 5232 Villigen-PSI, Switzerland

⊥ Laboratory of Inorganic Chemistry, Department of Chemistry and Applied Biosciences, ETH Zurich, 8093 Zurich, Switzerland

● , *Supporting Information*

## ABSTRACT

Hematite as promising photoanode for solar water splitting suffers from severe bulk and surface charge recombination. This work describes that a protonation-annealing treatment can effectively suppress both bulk and surface charge recombination in hematite. Protons/electrons are electrochemically incorporated into hematite under 0.2 $V_{RHE}$ followed by annealing at 120 $^{o}$C. The photocurrent density increases from ~0.9 mA cm$^{-2}$ to 1.8 mA cm$^{-2}$ at 1.23 $V_{RHE}$ under 1 sun, and further to 2.7 mA cm$^{-2}$ after loading cobalt phosphate, stabilizing at round 2.4 mA cm$^{-2}$. A cathodic shift of the onset potential of photocurrent is also observed. $H_2O_2$ oxidation, impedance spectroscopy and Mott-Schottky measurements show that the protonation suppresses bulk recombination and enhances donor density, but introducing more surface recombination. The annealing reduces surface recombination, while preserving relatively high bulk charge separation efficiency. Different from previous reports on the electrochemically reduced hematite, this work demonstrates that the performance improvement should be ascribed to the proton incorporation instead of the formation of $Fe_3O_4$ or metal Fe. This facile treatment by protonation and annealing could be applied in other semiconductors to promote the development of high performing photoelectrodes.

## 1. INTRODUCTION

The photoelectrochemical (PEC) splitting of water into hydrogen and oxygen is an attractive approach for converting solar energy into chemical fuels.[1-2] Photons absorbed by a semiconductor create electron-hole pairs, which can be separated by the space charge layer of the semiconductor to reduce/oxidize water into hydrogen and oxygen at the electrode-electrolyte interface. Hematite ($\alpha$-$Fe_2O_3$) has been intensively examined as photoanode for PEC water splitting[3-12] due to its theoretically high solar-to-hydrogen conversion efficiency (14−17%, corresponding to a photocurrent density of 11−14 mA cm$^{-2}$ at 1.23 V$_{RHE}$ under 100 mW cm$^{-2}$ irridation).[13-15] However, the reported photocurrent densities for water splitting using hematite photoanodes are far below the theoretical maximum value.[4-5, 7, 16-19] Only the electron-hole pairs which are photo-generated within the space charge layer or those can diffuse into this layer will be separated, while the rest of the electron-hole pairs will recombine in the bulk.[8, 13] Moreover, severe surface recombination has also been reported due to the photo-induced high valent iron species at the surface.[20-21]

Various morphologies of hematite photoanodes on fluorine-doped tin oxide (FTO) substrate have been reported including hydrothermally prepared nanorods,[4-5, 10, 16-18] solution-based colloids,[3] chemical vapor deposited nanostructures,[7, 22-25] spray pyrolyzed platelets[23] and deposition-annealing prepared nanoparticles.[19] According to literature,[4-5, 16-18] hydrothermally prepared hematite nanorods need to be activated by a short time annealing at about 800 ºC to enable tin diffusion into hematite from FTO substrate for higher donor density and better interface between hematite and FTO. Photocurrent densities[4-5, 17-18] with a large dispersion have been reported in the range of 0.6−1.3 mA cm$^{-2}$ for hematite nanorod photoanodes at 1.23 V$_{RHE}$, while loading cocatalysts NiFeO$_x$[4] or FeOOH[5] improves the photocurrent density to ~1.4 mA cm$^{-2}$. Doping with Zr[16], Pt[18] and P[26] results in higher photocurrent density of ~1.5 mA cm$^{-2}$, ~2.2 mA cm$^{-2}$ and ~2.7 mA cm$^{-2}$, which can be further increased up to over 3 mA cm$^{-2}$ after loading cobalt phosphate (Co-Pi) cocatalyst. An extra passivation layer of SiO$_x$[27], TiO$_2$[28] or Fe$_2$TiO$_5$[29] deposited by atomic layer deposition also demonstrates significant improvement in the PEC performance of hematite photoanode.

A negative polarization of metal oxide electrodes in aqueous electrolyte can simultaneously introduce electrons and small cations such as H$^+$ and Li$^+$ in the bulk structure, which is termed electrochemical doping.[30-36] The incorporation of proton/e$^-$ was found to improve the electrode efficiency for dye-sensitized solar cells[32-33] and water splitting.[30, 34-35] Proton functions as a shallow donor in semiconductors and increases the donor density, thus improving the majority carrier transport, as demonstrated for semiconductors with low donor densities like $WO_3$,[30] $TiO_2$,[32-33] $ZnO$,[37] $In_2O_3$ and $SnO_2$,[38] etc. Moreover, another reason for the improved photocurrent has been recently clarified that electrochemically incorporated protons readily segregates to the grain boundary reducing 50% of the overall number of deep electron traps.[31] The protonation through electrochemical doping is usually long lasting and can only be reversed by extremely long time positive polarization[31] or heating at high temperature.[30] On the other hand, a negative enough polarization can also reduce hematite to more conductive phases such as $Fe_3O_4$ or even Fe to enhance the performances.[34, 39-40]

This study aims at understanding the influence of electrochemical doping on the PEC performance of hematite nanorods photoanode prepared by hydrothermal method.[4-5, 16-18, 41] Protons are electrochemically introduced to hematite, followed by annealing treatments at an optimized temperature of 120 $^o$C. Indeed, the protonation and annealing processed hematite shows a photocurrent density of 1.8 mA cm$^{-2}$ at 1.23 $V_{RHE}$, which is higher than the values reported in literature for hematite nanorods photoanode without cocatalysts.[4-5, 16-17] After loading Co-Pi cocatalyst, hematite exhibits a photocurrent density of 2.7 mA cm$^{-2}$ at 1.23 $V_{RHE}$, which stabilizes at around 2.4 mA cm$^{-2}$ after 5 h test, comparable to the performances previously reported for hematite with different morphologies such as Ti-doped hematite nanoparticles,[19] $IrO_2$-loaded hematite cauliflowers.[7] XPS, Raman and Mössbauer spectra excludes the presence of $Fe^{2+}$ or Fe. Thus we ascribe the photocurrent change to the incorporation of proton/e$^-$, which is similar to the Volmer reaction for hydrogen evolution reaction. We clarify that the protonation improves the bulk charge separation efficiency but introduces more surface recombination, while the annealing treatment suppresses the surface recombination, still preserving relatively high bulk charge separation efficiency.

## 2. EXPERIMENTAL SECTION

**2.1. Preparation of hematite photoanode.** Fluorine-doped tin oxide (FTO) glass substrates (6~9 $\Omega$ sq$^{-1}$, Xop Glass) were cleaned by deionized water, isopropanol and acetone. $\beta$-FeOOH nanorods were grown on FTO in a solution containing 0.15 M FeCl$_3$ (97%, Aldrich) and 1 M NaNO$_3$ (99.5%, Merck) with a pH 1.5 adjusted by HCl. The reaction was carried out at 100 $^o$C for 6 h. Kapton tape was used to define the area for hematite growth and isolate the back side of the glass from the growing solution. After removing Kapton tapes and cleaning with DI water, FeOOH was annealed in air at 550 $^o$C for 2 h and then activated by annealing at 800 $^o$C for 10 min in a tube furnace to enable the diffusion of Sn from fluorine-doped SnO$_2$ (FTO) substrate to hematite.[4-5, 16-18]

**2.2. Co-Pi deposition**. The layer of Co-Pi on annealed hematite was loaded by photo-assisted electrodepostion under AM 1.5 G illumination from an electrolyte containing 0.5 mM Co(NO$_3$)$_2$·6H$_2$O (98%, Alfa Aesar) in 0.1 M potassium phosphate buffer at pH 7. A galvanostatic deposition was applied for 100 s with a current density of 6 $\mu$A cm$^{-2}$.

**2.3. Photoelectrochemical characterization.** All photoelectrochemical measurements were performed using a potentiostat/galvanostat Solartron 1286 electrochemical interface. The light source was a 150 W Xenon arc lamp equipped with an AM 1.5 G filter (100 mW cm$^{-2}$, Newport 66477-150XF-R1) calibrated by a photodetector (Gentec-EO). PEC measurements were performed in a three electrode configuration in a quartz cell in Ar-saturated 0.5 M NaOH (pH=13.0) electrolyte. Hematite/FTO photoelectrode was used as the working electrode, a Pt mesh and Ag/AgCl were used as the counter and the reference electrodes, respectively. Potentiodynamic measurements with a scan rate of 10 mV s$^{-1}$ in the potential window of 0.4–1.6 V$_{RHE}$ were performed to probe the photocurrents.

Open-circuit potentials were measured in O$_2$-saturated 0.5 M NaOH electrolyte, and each dark/light potential reading was obtained after stabilization for at least 30 min with oxygen gas bubbling into the electrolyte.

**2.4. Impedance spectroscopy and Mott-Schottky plots.** Dark and light impedance spectroscopy measurements were carried out in O$_2$-saturated 0.5 M NaOH electrolyte on a Solartron 1260 impedance analyzer combined with potentiostat Solartron 1286 in the frequency range between 10 mHz and 10 kHz using 5 mV AC amplitude. Mott-Schottky plots were obtained by fitting the dark impedance spectroscopy data. The sampling was taken every 50 mV in the potential region of 0.78-1.18 V$_{RHE}$, and prior to impedance measurement, hematite was stabilized at the measuring potential for 30 min. Light impedance spectroscopy was carried out at 1.18 V$_{RHE}$ and 0.98 V$_{RHE}$ under AM 1.5 G illumination. Data analysis was performed with ZView software.

**2.5. Materials characterizations.** The surface morphology of hematite was examined with a scanning electron microscope (SEM, Ultra 55, Carl Zeiss) operated at an accelerating voltage of 10 kV. X-ray photoelectron spectroscopy (XPS) measurements were done using a VG ESCALAB 220iXL spectrometer (Thermo Fischer Scientific) equipped with an Al Kα monochromatic source and a magnetic lens system. Mössbauer spectroscopy was done with conversion electron Mössbauer spectroscopy mode (CEMS) with electrons of about 5-10 keV. Raman spectra were recorded on Labram confocal Raman spectroscopy system, Jobin Yvon. X-ray diffraction (XRD) measurements were conducted by a Bruker−Siemens D500 X-ray Diffractometer. AFM was conducted on Agilent AFM5500 scanning probe microscope.

## 3. RESULTS AND DISCUSSION

**3.1. PEC performance of hematite.** Hematite nanorods on FTO were obtained by annealing the hydrothermally grown β-FeOOH nanorods in air at 550 °C for 2 h and then at 800 °C for 10 min. The as-prepared hematite shows a photocurrent density of ~0.9 mA cm$^{-2}$ at 1.23 V$_{RHE}$, and ~1.6 mA cm$^{-2}$ at 1.6 V$_{RHE}$, which are similar to previous reports in Ref. 4-5, 16.

The same hematite was then polarized at a slightly negative potential of 0.2 V$_{RHE}$ (−0.78 V *vs.* Ag/AgCl) for 200 s to incorporate proton/e$^{-}$ couples (protonation). After the 1$^{st}$ protonation, at highly positive potentials the hematite shows higher photocurrent densities than the as-prepared one (Figure 1a), but has lower photocurrent densities than the as-prepared one at moderately positive potentials between ~1.0 V$_{RHE}$ and ~1.35 V$_{RHE}$.

After protonation, hematite was annealed in air under a temperature of 120 °C for 1 h. Temperature and time were optimized by a series of annealing test under various temperature from 80 °C to 250 °C for different time (see Supporting Information Figure S1). This annealing treatment can improve photocurrent densities throughout the whole scanned potential region. It is worth mentioning that no color changes were observed between the as-prepared, protonated and annealed hematite, which has reddish orange color as shown in Figure S2a.

After the 2$^{nd}$ protonation process, the photocurrent density curve of hematite shows similar features as that of the 1$^{st}$ protonated hematite. Only under highly positive potentials, the 2$^{nd}$ protonated hematite has higher photocurrent densities than the 1$^{st}$ annealed hematite. Again, the $2^{nd}$ annealed hematite shows a further overall improvement throughout the whole scanned potentials compared to the $1^{st}$ annealed hematite. After these two rounds of protonation/annealing treatment, hematite has a photocurrent density of 1.8 mA cm$^{-2}$ at 1.23 V$_{RHE}$, twice of that for as-prepared hematite. In addition, the photocurrent onset potential is negatively shifted by ~80 mV in comparison to the as-prepared hematite. It is also noteworthy that the $2^{nd}$ protonated hematite exhibits improved photocurrent densities than the $1^{st}$ protonated hematite and the as-prepared sample. It is most likely that the relatively low temperature annealing process did not remove all the introduced protons, instead some of the protons are resistant to the annealing treatment. The $3^{rd}$ protonated hematite exhibits better performance than the $2^{nd}$ protonated one, however, the following $3^{rd}$ annealing treatment does not obviously improve the performance of hematite.

A layer of Co-Pi cocatalyst was photo-electrodeposited on the processed hematite, and the photocurrent density is also shown in Figure 1a. Co-Pi decorated-hematite has a photocurrent density of ~2.7 mA cm$^{-2}$ at 1.23 V$_{RHE}$, and ~3.6 mA cm$^{-2}$ at 1.6 V$_{RHE}$, which is comparable to the benchmark performances previously reported for hematite with different morphologies such as Ti-doped hematite nanoparticles,[19] IrO$_2$-loaded hematite cauliflowers.[7] The stability test show that the Co-Pi decorated-hematite has a photocurrent density of ~2.4 mA cm$^{-2}$ at 1.23 V$_{RHE}$ after 5 h test (see Figure 1b).

Scanning electron microscopy (SEM) characterization was used to investigate the morphology changes of the as-prepared and treated hematite samples, which were obtained by cutting a single hematite followed by different treatments. Figure 1c-e show the typical features of hematite nanorods, which are similar to previous reports in literature.[18, 40, 42] Two typical features of nanorods are observed. One type is round and vertical to the substrate, while the other is more elongated and tilted to the substrate. The amount ratios between the two nanorods for these three samples seem to be not uniform according to the SEM and AFM images (Figure S5), however, a meaningful correlation between this phenomenon and the protonation/annealing treatment cannot be directly made. It is more likely that this discrepancy forms due to the collapse, shrink, and coalescence of β-FeOOH precursors during the high temperature sintering process, which is also one of the reasons that the photocurrents of the as-prepared hematite are not always the same but rather fluctuates in a range. In our experiments, the photocurrent densities of as-prepared hematite at 1.23 $V_{RHE}$ are found in the range of 0.9-1.2 mA cm$^{-2}$. In literature, photocurrent densities[4-5, 17-18] with an even larger dispersion have also been reported between 0.6−1.3 mA cm$^{-2}$ for hematite nanorod photoanodes at 1.23 $V_{RHE}$.

To clarify the effects of protonation and annealing treatment, a control sample was annealed at 120 ºC for 1 h without beforehand protonation (Figure S2b). The annealing process alone leads to a slight improvement in the photocurrent densities only at very high potentials, and the onset potential remains the same. In the second round treatment, both protonation and annealing were applied, and then a smaller onset potential and higher photocurrent are observed.

The change tendencies of photocurrents after protonation and annealing treatment are consistent for different samples, in spite of the fluctuated initial photocurrents, as shown in Figure 1, S1 and S2. Therefore, it would still be plausible to conclude that this protonation-annealing treatment is an effective strategy to enhance the performances for hematite photoanode.

**3.2. Photocurrents of water and H$_2$O$_2$ oxidation.** To understand the effects of protonation and annealing treatment on hematite, H$_2$O$_2$ was used as hole scavenger in electrolyte. The oxidation of H$_2$O$_2$ is thermodynamically more facile than oxidation of water, and has a rate constant 10-100 times higher than that of water.[23] Thus the oxidation of H$_2$O$_2$ is considered as a process with minimal losses of holes at the surface, reflecting mainly the bulk charge separation properties of hematite. A smaller gap between the oxidation photocurrents of $H_2O_2$ and $H_2O$ indicates less surface recombination of charge carriers.

It is noteworthy that the same hematite (before loading Co-Pi) as in Figure 1 was used to probe the photocurrent densities for $H_2O_2$ oxidation. Figure 2a-c show the comparisons between the photocurrent densities of $H_2O_2$ and water oxidation for as-prepared, protonated and annealed hematite. The gap between the photocurrent densities of $H_2O_2$ oxidation and $H_2O$ oxidation increases in the following order: annealed < as-prepared < protonated, indicating that the protonated sample suffers from the most severe losses of holes at the surface or close to the surface, while the annealed hematite has the lowest surface recombination.

The bulk charge efficiency can be obtained by dividing the photocurrent $J_{H2O2}$ over $J_{abs}$ (absorption derived maximum photocurrent) assuming the charge injection efficiency is 100% in the presence of $H_2O_2$.[23] $J_{abs}$ was calculated by integrating the light absorption of each sample with respect to AM 1.5G spectrum. The bulk charge separation efficiency is thus calculated and plotted in Figure 2d. Clearly, the protonated and annealed hematite both have enhanced charge separation efficiency in the bulk, while the former has the highest value, most likely due to higher majority carrier transport and lower bulk recombination. This reduced bulk recombination has been observed on rutile $TiO_2$ nanoparticle electrodes,[31] and explained by the reduction of deep electron traps at the grain boundary as a result of proton incorporation.

To summarize, the protonation improves the bulk charge separation efficiency but introduces more surface recombination. The annealing treatment can reduce the surface recombination at the sacrifice of certain bulk properties, although it still has higher charge separation efficiency than the as-prepared hematite.

**3.3. Impedance spectroscopy measurements under light.** Impedance spectroscopy were measured under AM 1.5G illumination (100 mW cm$^{-2}$) to further understand the effects of protonation and annealing on the bulk electron transport and solid-electrolyte interface charge transfer processes. Figure 3a shows the Nyquist plots acquired at 1.18 V$_{RHE}$ under illumination for as-prepared and processed hematite. The equivalent circuit[8, 13] shown in the inset of Figure 3a was used to fit the data, which includes both the bulk electron transport and solid-electrolyte charge transfer processes. Here the bulk includes both space charge and Helmholtz layers. Figure 3b reports the fitted resistances and time constants for bulk electron transport and interface charge transfer at 1.18 V$_{RHE}$.

The protonated hematite has the smallest bulk resistance and time constant, but the largest charge transfer resistance and time constant. This is consistent with the conclusion derived from the analysis of the photocurrent measurements for water and H$_2$O$_2$ oxidation that the protonation improves the bulk charge separation efficiency but also introduces more surface recombination. Water oxidation on metal oxide electrodes is believed to proceed from surface hydroxyl intermediate states formed by hole transfer to a surface-coordinated water and a concomitant deprotonation.[20, 43-45] The extra protons near the surface of the protonated hematite probably inhibit the hole transfer to water, which results in large resistance R$_{ct}$ and time constant (RC)$_{ct}$ for charge transfer process.

The annealed hematite has the smallest charge transfer resistance and time constant, as well as the second smallest bulk resistance and time constant, which is also in good agreement with the photocurrent measurements. It is most likely that the annealing process removes mainly the protons near the surface, largely suppressing the surface recombination while still preserving relatively high bulk charge separation efficiency.

Impedance spectra were also collected at 0.98 $V_{RHE}$ (Figure S3a) and the fitted parameters are shown in Figure S3b. The protonated hematite has the largest time constants for both the bulk transport and charge transfer processes, which is consistent with its very low photocurrents at low potentials.

**3.4. Mott-Schottky plots.** As previously mentioned, some protons may be resistant to the annealing process. If this is true, the donor density of the annealed hematite should be higher than that of as-prepared one. Impedance spectroscopy data measured in dark were fitted according to the equivalent circuit in the inset of Figure 4a, which has been proved to best represent the dark phenomena.[8] The fitted space charge layer capacitances were used to plot Mott-Schottky curves (Figure 4a), which enables us to obtain the donor density and flat band potential. The details for fitting impedance spectra, calculating donor density and flat band are shown in the Supporting Information.

The donor densities obtained from Mott-Schottky plots are shown in Figure 4b. The as-prepared hematite has a donor density of $1.54 \times 10^{19}$ cm$^{-3}$, which is comparable to literature reports on Sn-doped hematite[17] and Si-doped hematite.[8] Repeated protonation progressively increases the donor density, which rises up to $4.48 \times 10^{19}$ cm$^{-3}$ after the 3$^{rd}$ protonation, contributing to the high conductivity and charge separation efficiency in bulk for protonated hematite. The 3$^{rd}$ annealed hematite still has higher donor densities than the as-prepared hematite. This confirms that residual protons are present in the annealed hematite, probably due to higher removal energy barriers of the protons in the bulk than those near the surface.

It is noteworthy that the flat band potential slightly shifts to more positive values (downward shift) upon repeated protonation and annealing (Figure 4b). According to eq 1,[46-47] the width of the space charge layer $W_{sc}$ under a certain potential $V$ can be obtained when the flat band potential $V_{fb}$ and donor density $N_D$ are known:

$$W_{sc} = \left( \frac{2(V - V_{fb})\varepsilon\varepsilon_0}{eN_D} \right)^{1/2} \tag{1}$$

The calculated $W_{sc}$ for different samples at 1.23 $V_{RHE}$ are listed in Figure 4b. $W_{sc}$ decreases upon protonation, and increases upon annealing. A thin space charge layer cannot effectively block the back recombination of surface accumulated holes with bulk electrons.[48] This explains the relatively low photocurrents for water oxidation of the protonated sample at moderately anodic potentials. While at highly positive potentials, the space charge layer becomes wide enough to block the back recombination, leading to similar or higher photocurrents than those measured for the as-prepared hematite.

The influence of oxygen vacancies on the donor density can be ruled out since oxygen vacancy in hematite has very high formation energy of ~20-23 eV,[49-50] which makes the formation of extra oxygen vacancy almost impossible at 120 °C. Besides, the minor change in photovoltage (Figure S4) cannot be the main reason for enhanced PEC performance.

**3.5. Mechanism discussion.** The protonation treatment was conducted under 0.2 $V_{RHE}$ (-0.78 V *vs*. Ag/AgCl) in 0.5 M NaOH in our experiment. According to the Pourbaix diagram of Fe-H$_2$O system,[51] this potential still corresponds to the region of $Fe_2O_3(s)$. Previous reports on the electrochemical reduction of hematite[34, 39-40] often ascribe the enhancement of photocurrents to the formation of more conductive $Fe_3O_4$ or even metal Fe. However, more negative potentials (between -1 V and -1.5 V *vs.* Ag/AgCl or SCE.) were used in these works to reduce hematite, which correspond to the region of $Fe_3O_4$ or even more reduced phases in Pourbaix diagram. In addition, Cibrev *et al.*[40] have elaborated the electronic structure changes of hematite after electrochemical reduction through photoemission (PES) and X-ray absorption (XAS) and demonstrated that only potentials more negative than -1.2 V *vs.* Ag/AgCl are enough to introduce new electronic states in the vicinity of valance band of hematite.

In this work, however, no evidence supports the presence of $Fe_3O_4$ or more reduced phases of iron. Figure 5 shows XPS, Raman and Mössbauer spectra for as-prepared and treated hematite. No obvious differences are observed between the as-prepared, protonated and annealed hematite for all the spectra. XPS spectra show that binding energies of Fe $2p_{3/2}$ and Fe $2p_{1/2}$ are 710.3 eV and 723.6 eV, respectively, along with a typical satellite peak at 718.5 eV, which is characteristic of $Fe^{3+}$.[40, 52] However, no shoulder close to the Fe $2p_{3/2}$ peak corresponding to the presence of $Fe^{2+}$ was detected after the protonation or annealing treatments. Raman and Mössbauer spectra are also virtually the same for as-prepared, protonated and annealed hematite, only showing the characteristics of $Fe_2O_3$. Besides, XRD patterns in Figure S6 also confirm that crystallographic structures are identical for as-prepared and treated hematite. All these evidences verify that our electrochemical reduction do not involve the formation of magnetite $Fe_3O_4$ or metal Fe.

Based on the above discussions, it would be justified to draw the conclusion that the enhancement of photocurrents in this work is related with protons instead of the reduction of $Fe_2O_3$.

A new mechanism is proposed for this work to explain the protonation process of hematite, which can be written similar to the Volmer equation:[53-54]

$$Fe_2O_3 + H_2O + e^- = Fe_2O_3H_{ads/int} + OH^- \qquad (2)$$

Protons can be adsorbed onto or intercalated into hematite, which simultaneously introduces electrons into $Fe_2O_3$ and generates $OH^-$ ions in the local region. Here electrons increase the donor density in the bulk of hematite, which is reflected by the improved charge separation efficiency and conductivity in bulk for protonated hematite. But protons near the surface may also act as recombination centers thus deteriorate the photocurrent under relatively low bias. It is found that a low temperature annealing treatment can recover the surface properties. This has never been reported in literature. We thus proposed a combined protonation-annealing treatment for hematite. After annealing treatment, extra protons on the surface were removed and thus the surface recombination was also impressed.

Furthermore, previous reports on electrochemical reduction of hematite used sample prepared under a relatively low temperature of 550 °C[34] or 600 °C[39-40], which have very low photoactivity. For hematite with low donor densities, the electrochemical reduction of hematite can improve the photocurrents for water splitting through the enhancement of the majority carrier transport. Still, photocurrent densities of less than 1 mA cm$^{-2}$ are reported for optimized pretreated hematite at 1.23 $V_{RHE}$ in these works.[34, 39-40] In our work, however, an additional activation treatment by sintering hematite at 800 °C for 10 min was adopted to enable the diffusion of Sn from fluorine-doped $SnO_2$ (FTO) substrate to hematite.[4-5, 16-18] Therefore, this work used highly-doped hematite, which is also verified by the much higher photocurrents in our case. Although Shangguan et al.[34] also reported a Ti-doped hematite, the annealing treatment under 550 °C seems inefficient to incorporate enough $Ti^{4+}$ in the crystalline structure of hematite, which shows a very low photocurrent of ~400 µA cm$^{-2}$ at 1.23 $V_{RHE}$ in comparison to 1.8 mA cm$^{-2}$ in our case. Therefore, it is worth nothing that our new understanding on the protonation treatment is based on hematite with sufficiently high donor density.

## 4. CONCLUSIONS

In conclusion, repeated protonation and annealing treatments can cathodically shift the onset potential and improve photocurrents for hematite photoanode. Literature often ascribe the enhancement in performance to the formation of $Fe_3O_4$ or metal Fe, however, it is demonstrated that the enhancement is mainly related with the incorporation of proton/e$^-$ in this work. Protons and electrons are incorporated into hematite in a way similar to Volmer reaction. The protonation process increases the donor density, reduces bulk electron traps but introduces more surface recombination. The subsequent low temperature annealing removes extra protons near the surface to suppress surface recombination while preserving relatively high bulk charge separation efficiency. For oxides with low donor densities, a negative polarization alone can improve the photocurrents for water splitting through the enhancement of the majority carrier transport, as in the case of low-doped $Fe_2O_3$.[34, 39-40] However, for oxides with a sufficiently high donor density, further increasing the donor density will reduce the photocurrents due to the increased recombination near the surface. Our measurements show that a low temperature annealing treatment can improve the surface properties while keeping a low bulk recombination.

■ **ASSOCIATED CONTENT**

●, **Supporting Information**

The Supporting Information is available free of charge on the ACS Publications website at DOI: 10.1021/acsami.xxxxxx.

Additional experimental details including optimization of annealing parameters, photograph of hematite, photocurrents for anneal-only hematite, Nyquist plot under light at 0.98 $V_{RHE}$, open circuit potentials, electrochemical surface area, photovoltage calculations, Pourbaix diagram, AFM images and XRD patterns. (PDF)

## ■ AUTHOR INFORMATION

**Corresponding Author**

* siwp86@gmail.com; thomas.lippert@psi.ch

**Notes**

The authors declare no competing financial interest.

## ■ ACKNOWLEDGEMENTS

This research is supported by the Paul Scherrer Institute and the NCCR MARVEL, funded by the Swiss National Science Foundation. We thank U. Aschauer for discussions of the results, A. Palla-Papavlu, M. El Kazzi, J. Landers and H. Wende for kind help with experiments.

## ■ REFERENCES

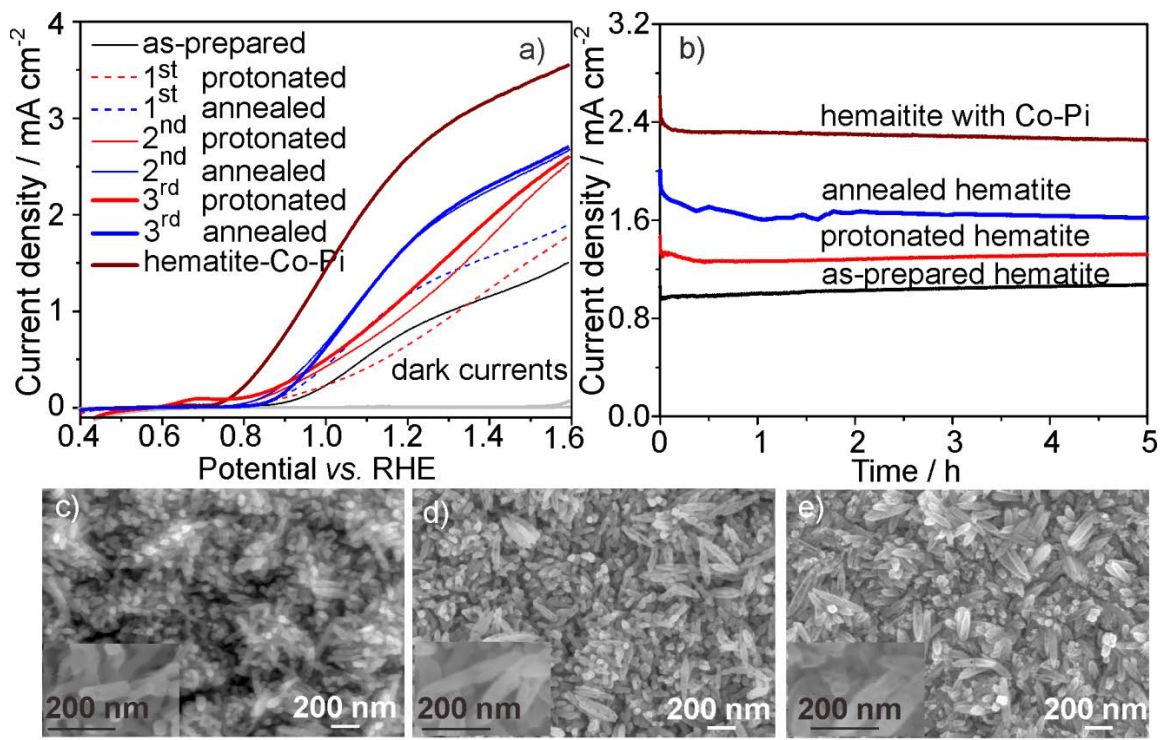

**Figure 1.** (a) Photocurrent densities of hematite in Ar-saturated 0.5 M NaOH electrolyte (pH=13.0) under AM 1.5G illumination of 100 mW cm$^{-2}$: as-prepared, the 1$^{st}$ protonated, the 1$^{st}$ annealed at 120 $^o$C, the 2$^{nd}$ protonated, and the 2$^{nd}$ annealed at 120 $^o$C, as well as processed hematite with Co-Pi cocatalyst. (b) Photocurrent stability test for protonated, annealed hematite and hematite with Co-Pi cocatalyst. (c-e) SEM images for as-prepared, protonated, and annealed hematite.

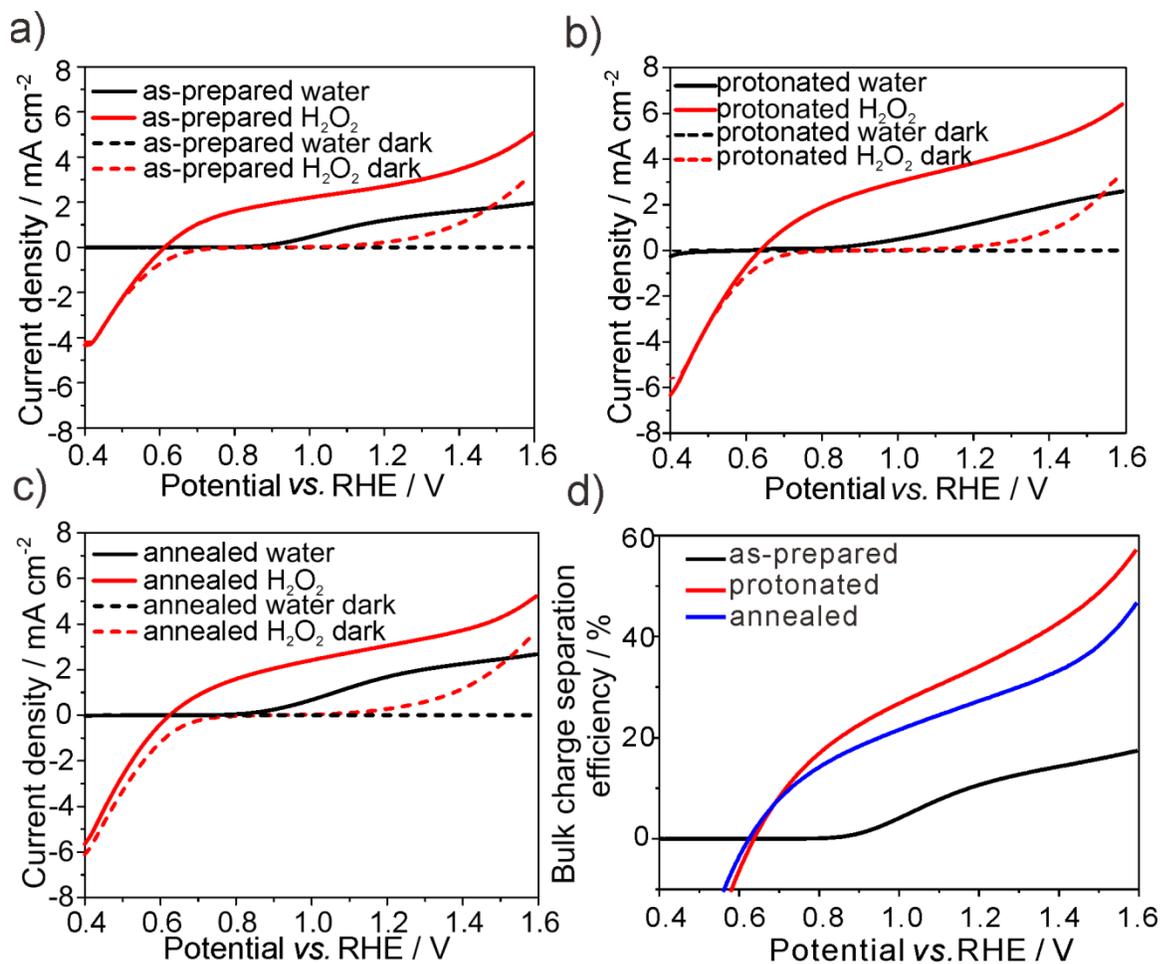

**Figure 2.** Photocurrent densities of hematite in 0.5M NaOH in the presence of 0.5 M H$_2$O$_2$ under AM 1.5G illumination with a power density of 100 mW cm$^{-2}$. (a) As-prepared, (b) the 3$^{rd}$ protonated, (c) annealed at 120 $^o$C after the 3$^{rd}$ protonation. (d) The bulk charge separation efficiency for as-prepare, protonated and annealed hematite.

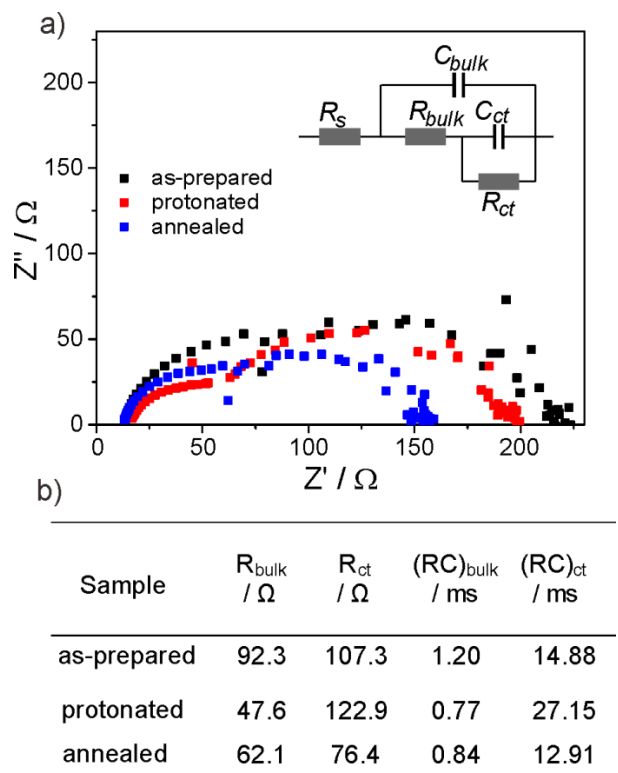

a)

b)

| Sample | $R_{bulk}$ / $\Omega$ | $R_{ct}$ / $\Omega$ | $(RC)_{bulk}$ / ms | $(RC)_{ct}$ / ms |
|---|---|---|---|---|
| as-prepared | 92.3 | 107.3 | 1.20 | 14.88 |
| protonated | 47.6 | 122.9 | 0.77 | 27.15 |
| annealed | 62.1 | 76.4 | 0.84 | 12.91 |

**Figure 3.** (a) Nyquist plots measured under AM 1.5G illumination at 1.18 $V_{RHE}$ for as-prepared, 3[rd] protonated and annealed hematite in 0.5 M NaOH electrolyte saturated with $O_2$ and fitted with the equivalent circuit shown as inset. (b) Bulk transport resistance $R_{bulk}$ and time constant $(RC)_{bulk}$, hematite-electrolyte charge transfer resistance $R_{ct}$ and time constant $(RC)_{ct}$, obtained by fitting the impedance spectroscopy data.

a)

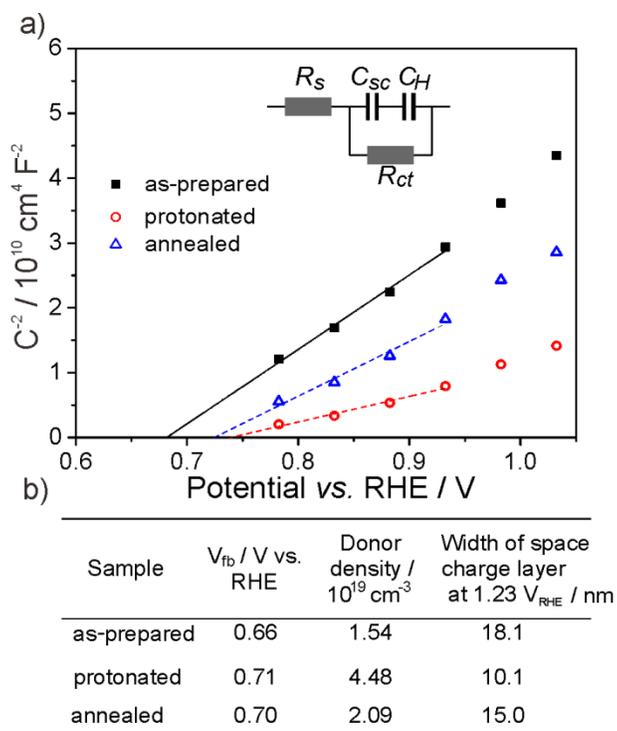

b)

| Sample | $V_{fb}$ / V vs. RHE | Donor density / $10^{19}$ cm$^{3}$ | Width of space charge layer at 1.23 $V_{RHE}$ / nm |
|---|---|---|---|
| as-prepared | 0.66 | 1.54 | 18.1 |
| protonated | 0.71 | 4.48 | 10.1 |
| annealed | 0.70 | 2.09 | 15.0 |

**Figure 4.** (a) Mott-Schottky plots obtained by measuring the impedance spectroscopy in 0.5 M NaOH electrolyte saturated with $O_2$ in dark for as-prepared, 3rd protonated and annealed hematite. The complex impedance plane plots were fitted according to the equivalent circuit in the inset. (b) Flat band potential, donor density and the width of space charge layer at 1.23 $V_{RHE}$.

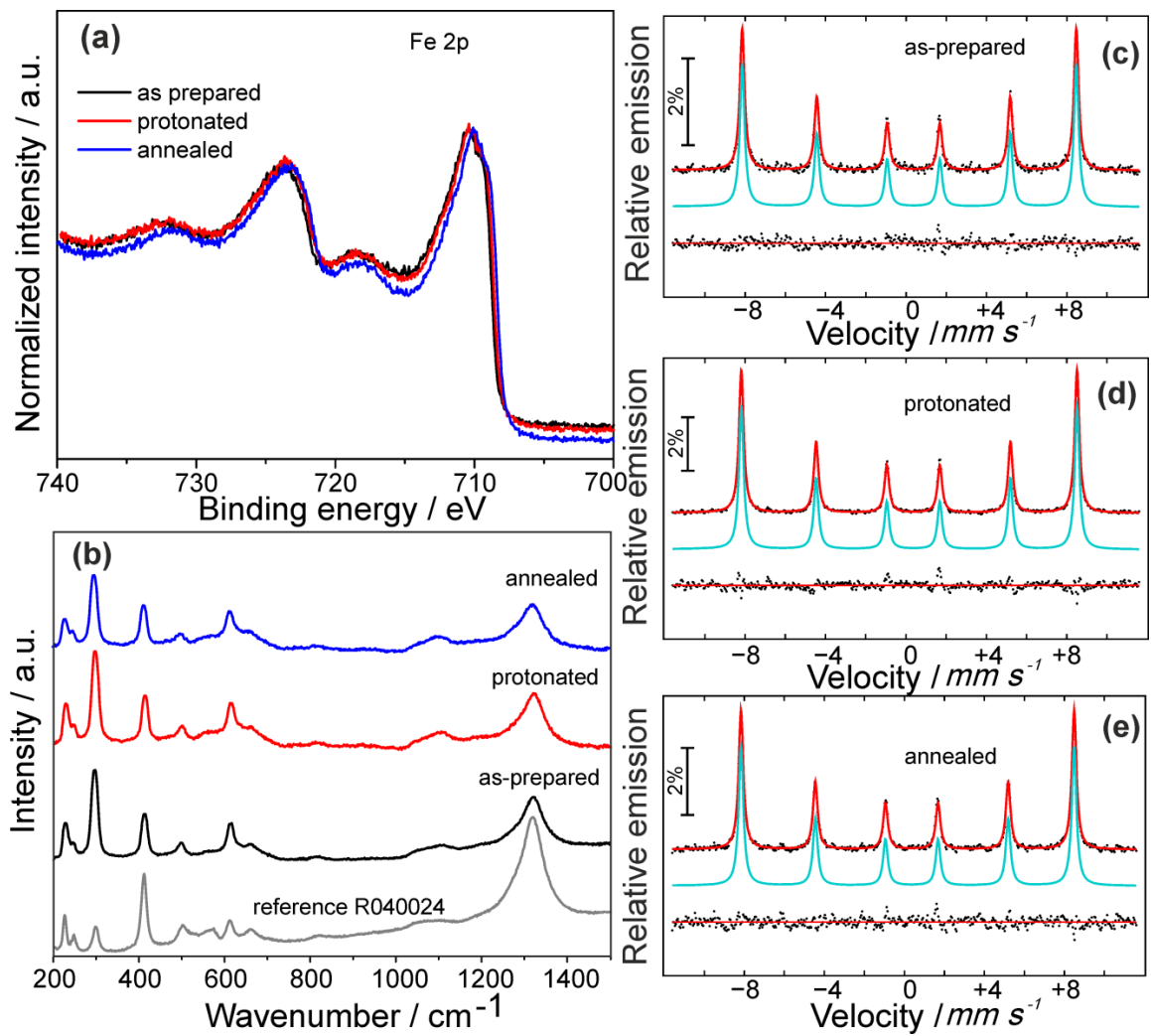

**Figure 5.** (a) X-ray photoelectron spectroscopy at the surface, (b) Raman and (c) Mössbauer spectroscopy in the bulk for as-prepared, protonated and annealed hematite.